\documentstyle[mprocl,epsfig]{article}
\def\be{\begin{equation}}
\def\ee{\end{equation}}
\def\bea{\begin{eqnarray}}
\def\eea{\end{eqnarray}}

\begin{document}
\title{SELF-ORGANIZATION OF COMPLEX SYSTEMS}
\author{MAYA PACZUSKI}
\address{Department of Physics, 
University of Houston,\\Houston TX 77204-5506, 
USA\\E-mail: maya@uh.edu}  
\author{PER BAK}
\address{Niels Bohr Institute, Blegdamsvej 17,
2100 Copenhagen, Denmark\\E-mail: bak@nbi.dk}

\maketitle \abstracts{The basic laws of physics are simple, so why is
the world complex? The theory of self-organized criticality posits
that complex behavior in nature emerges from the dynamics of extended,
dissipative systems that evolve through a sequence of meta-stable states
into a critical state, with long range spatial and temporal
correlations.  Minor disturbances lead to intermittent events of all
sizes.  These events organize the system into a complex state that cannot be
reduced to a few degrees of freedom.  This type of
``punctuated equilibrium'' dynamics
 has been observed in astrophysical,
geophysical, and biological processes, as well as in human social
activity.}

\section{Introduction}
Scientific inquiry in the second millennium has focused almost
exclusively on discovering the fundamental constituents, or building
blocks, of nature.  The most innermost secrets have been revealed down
to ever smaller scales.  Matter is formed of atoms; atoms
are composed of electrons, protons, and neutrons, and so on down
to the smallest scale of quarks and gluons. These basic elements
interact through simple physical laws.  

In the realm of biology, it is known that life on earth is based on
the DNA double helix.  But even though we understand perfectly the
laws governing the interaction of atoms, we cannot directly
extrapolate these laws to explain the beginning of life, or the
auto-catalysis of complex molecular networks, or why we have brains
that can contemplate the world around us.  Due to the overwhelming
unlikeliness of random events leading to complex systems like
ourselves, it seems as if an organizing agent or ``God'' must be
invoked who puts the building blocks together.

It isn't necessary to delve into the biological realm to see the
ultimate inadequacy of a purely reductionist approach.  For instance,
the surface of the earth is an intricate conglomerate of mountains,
oceans, islands, rivers, volcanoes, glaciers, and earthquake faults,
each with its own dynamics.  The behavior of systems like these cannot be
deduced by examining ever smaller scales to derive microscopic laws;
the dynamics and form is ``emergent.''  Unless one is willing to
invoke an organizing agent of some sort, all these phenomena must be
self-organized.  Complexity must emerge from a
self-organizing dynamics. But how?

A few ideas have been proposed that begin to address this problem,
which can be characterised as ``How do we take God out of the
equations.''  The most pessimistic view is that one has to
describe each and every feature in nature on a case by case basis.
Indeed, such a ``stamp collection'' approach has prevailed in sciences
such as biology and geophysics, and attempts to look for a unifying
description have in the past been met by very strong scepticism among
the practitioners of those sciences, although there have been
exceptions such as plate tectonics theory,  Kauffman's work
on autocatalytic networks \cite{Kauffman}, and Gould and Eldridge's theory of
punctuated equilibrium in biological evolution
\cite{Gould}.

Perhaps nature does not need to invent a multitude of mechanisms,
one for each system.  The view that only a limited number of
mechanisms, or principles, lead to complexity in all its manifestations
(from the galactic or universal to the molecular)
is supported by the observation of regularities that appear in the
statistical description of complex systems.  These statistical
regularities provide hope and encouragement that a science of
complexity may eventually emerge.

For example, river networks, mountain ranges, etc. exhibit scaling
behavior, both in the spatial and in the temporal domain, where
landslides or sediment deposits interrupt the quiet steady state.
These landslides have been observed to be scale free \cite{landslide};
similarly the Gutenburg Richter law for earthquakes states that they
are also a scale free phenomena, with avalanches (quakes) of all sizes
\cite{Richter}. The distribution of energy released during earthquakes
is a simple power law, despite the enormous complexity of the
underlying system, involving a multitude of geological structures.
Forest fires have a similar behavior \cite{Turcotte}, as does volcanic
activity \cite{volcano}.  In astrophysical phenomena, there are star
quakes, which we observe as pulsar glitches \cite{pulsar}, interrupting
quiet periods.  Black holes are surrounded by accretion disks, from
which the material collapses into the black hole in intermittent,
earthquake-like events, which interrupt the otherwise steady evolution
and occur over a wide range of scales \cite{blackhole}.

Biological evolution also exhibits long periods of stasis
punctuated by extinction events of all sizes.
The paleontologists Stephen Jay Gould and Niles Eldredge \cite{Gould}
coined the term ``punctuated equilibrium'' to describe the pace of
evolution.
Gould also argues that the
record of extinction of species is contingent on seemingly minor
accidents, and if the tape of the history of life were to be rerun
an entirely different set of species would emerge \cite{Gould2}.

We assert that  punctuated equilibrium dynamics is the 
essential dynamical process for
 everything that evolves and becomes complex, with a
specific behavior that is strongly contingent on its history
\cite{pnas}.  The periods of stasis allow the system to remember its
past, the punctuations allow change in response to accumulated forcing
over long time scales, and the criticality assures that even minor
perturbations can have dramatic effects on the specific outcome of a
particular system, making it possible to have distinct individual
histories and forms.

Perhaps the greatest challenge is to find the mechanism by which the
big bang has led to ever increasing complexity in our universe, rather
than exploding into a simple gas-like fragmented substance, as
explosions usually do, or imploding into a simple solid or black
hole. Some intricately balanced feature of the initial state must have
existed that allowed this to happen.  How that ``fine tuning'' could
have appeared remains a mystery, with Lee Smolin's speculation of
universes created by Darwinian selection being the only attempt so far
\cite{Smolin}.

Complexity is a hierarchical phenomenon, where each level
of complexity leads to the next: astrophysics, with its own
hierarchy of scales, leads to geophysics,
which is the prerequisite for chemistry, biology, and ultimately
the social sciences.  Although the origin of the hierarchy is
not understood, we do have the rudiments of a theory for the
emergence of one level out of the previous one.  Due to this hierarchy
of emergence, it isn't necessary to understand the mechanism of the
big bang in order to understand the dynamics of earthquakes.

A common feature of the systems mentioned thus far, and perhaps
of all complex systems, is that they are
 driven by slowly pumping in energy from a lower level of the
 hierarchy. For instance, biological life is driven by the input of
 energy from the sun. The energy is stored and later dissipated, in an
 avalanche process like an earthquake. Even a small increment in
 energy can trigger a large catastrophe, making these systems strongly
 contingent on previous history.  They operate far from equilibrium,
 which is necessary since systems in equilibrium tend to become more
 and more disordered (rather than complex) over time, according to the
 second law of thermodynamics.

\section{Complexity and Criticality}

One view of systems driven out of equilibrium is that they should tend
to a uniform ``minimally'' stable state generated by
some type of optimization process.  In traffic flow such a state would
correspond to a uniform flow of cars with all cars moving at maximum
velocity possible.  But these  optimized states often
are catastrophically unstable, exhibiting breakdown events or avalanches,
such as traffic jams \cite{np95}.  In tokamaks \cite{tokamak1},
this means that the ideal state of the plasma with the highest
possible energy density is locally stable, but globally unstable with
respect to explosive breakdown events. The surface of the sun is
unstable with respect to formation of solar flares emitting energy in
terms of light or gamma rays.  In fact, the actual sets of states that
emerge are those which are organized by the breakdown events. 

A possible self-organized state is one that is critical in the sense
that it has power law spatial and temporal correlations, like
equilibrium systems undergoing a second order phase transition. The
breakdown events in that state then must also be critical in the sense
of a nuclear chain reaction process. In a supercritical system, a
single local event, like the injection of a neutron, leads to an
exponentially exploding process. A sub-critical process has
exponentially decaying activity, always dying out.  In the critical
state, the activity is barely able to continue indefinitely, with a
power law distribution of stopping times, reflecting the power law
correlations in the system and {\it vice versa}.  

It is intuitively clear that complex systems must be situated at this
delicately balanced edge between order and disorder in a self-organized
critical (SOC) state.  In the ordered
state, every place looks like every other place. Think of a crystal
where the atoms are lined up over millions of inter-atomic
distances. In the disordered state, there are no correlations between
events that are separated in time or space: we have white
noise. Again, it makes no sense to talk about complex
behavior. Chaotic systems belong to this latter category. Sub-critical or
supercritical states can usually be understood quite easily by
analysing the local properties. Only at the critical state, 
does the compromise between order and surprise exist that
can qualify as truly complex behavior. There are very large
correlations, so the individual degrees of freedom cannot be
isolated. The infinity of degrees of freedom interacting with
one another cannot be reduced to a few.  This
irreducibility is what makes critical systems complex.

Thus, self-organized criticality provides a general mechanism for the
emergence of complex behavior in nature.  
It has been proposed that
granular piles \cite{BTW}, traffic \cite{np95}, 
magnetic fusion plasmas \cite{tokamak1},
the crust of the earth \cite{bak-tang,ito}, river networks
\cite{rinaldo} and braided rivers \cite{minnesota,bassler2},
superconductors in a magnetic field \cite{field}, etc.,
all operate in a self-organized critical state.

The sandpile was the first model introduced by Bak, Tang, and
Wiesenfeld to demonstrate the principle of self-organized criticality
\cite{BTW,B}.  This  model has subsequently received a great
deal of attention due in part to its potential for having a
theoretical solution.  Dhar showed that certain aspects of its
behavior could be calculated exactly based on the Abelian symmetry of
topplings \cite{Dhar1}.  The sandpile was thought of as a paradigmatic
gedanken experiment, but there has also been experimental
confirmation of self organized criticality in granular
piles. Fig.~1 shows an experiment on a pile of rice by Frette et
al. \cite{frette}.  Grains of rice were dropped between two glass
plates by a seeding machine, and the avalanches were monitored by a
video camera connected with a computer for data analysis. A power law
distribution of avalanches was found, indicating SOC.

\begin{figure}[t] 
\vspace{10cm}     
 \begin{center}
    \leavevmode   
\caption[]{Avalanche in Ricepile Experiment}
  \end{center}
\end{figure} 

Over the past decade there has been a great deal of theoretical work on
other models of SOC. Much of this work has focussed
on other idealized models of sandpiles.  These models typically involve a
sequence of nodes to which sand is added until a critical gradient or
height is reached locally, triggering redistribution of sand to
nearest neighbors.  Then a chain reaction of instabilities may occur
encompassing all scales up to the system size.  Self-organized
critical  systems  evolve toward a scale-free, or
critical state naturally, without fine tuning any parameters.  This
gives rise to power law distributions for the breakdown events.
Minimal SOC models have been developed to
describe a diverse set of phenomena including earthquakes 
\cite{bak-tang,ito,sornette,carlson-langer,KimC},
solar flares \cite{Lu}, forest fires \cite{forestfire}, 
magnetically confined plasma 
\cite{tokamak1,tokamak2}, fluctuations in stock-markets 
\cite{bps} and economics
\cite{BakSch},
black hole accretion disks \cite{blackhole,blackhole2}, 
traffic \cite{np95}, biological evolution
\cite{BS,pmb}, braided rivers formed by
vortex avalanches in superconductors \cite{kevin}, and disease epidemics
\cite{RA}, among others \cite{B}.

Given the preliminary nature of current  understanding of
complex systems, we are forced to consider one type of system at a
time, looking for general principles.  Some advancement has come
from developing and studying simple computer models which help to
conceptualize the essential attributes of
the specific phenomena, and eventually to relate those to
other phenomena.  In the following we shall review a couple of these
applications from widely different scientific domains: one from biology
(co-evolution of species), one from solid state  and geophysics 
(vortex avalanches and braided rivers), one from the social sciences (traffic),
and one from cognitive science (brain function).

\section{Braided Rivers  and Superconducting Vortex Avalanches}

Magnetic flux penetrates type II superconductors in quantized vortices
which can move when an electrical current is applied, overcoming
pinning barriers. When magnetic flux is forced in or out of the
superconductor, vortices have been observed to intermittently flow
\cite{field} through preferred channels \cite{japan}.  Using a simple
cellular model \cite{kevin} to mimick this experimental situation,
it has been found that the vortex flow makes rivers strikingly similar
to aerial photographs of braided fluvial rivers, such as the
Brahmaputra \cite{braidedrivers}. This suggests that a common
dynamical mechanism exists for braiding, namely, avalanches of
stick-slip events, either sliding sediment or vortices, which organize
the system into a critical braided state \cite{bassler2}.

The cellular model \cite{kevin}
includes basic features of vortex dynamics: over-damped motion of
vortices, repulsive interactions between vortices, and attractive
pinning interactions at defects in the material.   It is a coarse
grained description at the scale of the range of intervortex interactions,
the so-called London length, and throws out most microscopic degrees
of freedom (specific information about the vortex cores). As in experiments,
vortices are slowly pushed into the system at one boundary (the left)
and allowed to leave at the other boundary (the right).  The
vortex-vortex repulsions cause a gradient to build up in the vortex
density across the system. Eventually, as vortices are constantly
added, a critical slope is achieved where the force from the gradient
of vortex density is opposed by pinning forces, making a delicately
balanced vortex pile reminiscent of a pile of sand.  Then adding new
vortices slowly at the boundary triggers avalanches of vortex motion,
where one moving vortex can cause others to become unstuck, leading to
a chain reaction.  Avalanches of all sizes occur, limited only by the
physical size of the system.  Since the avalanches have no other
characteristic spatial or temporal scale, the model exhibits
self-organized criticality. Similar behavior has been
observed experimentally\cite{field}, and in molecular
dynamics simulations of the microscopic equations of motion\cite{olson}.

The spatial variation of the overall vortex flow is measured in terms
of the number of vortices moving in each cell, averaged over a long
time interval representing many vortices flowing through the system.
Fig.~2.  represents a ``time-lapsed'' photograph of vortex motion.
Rather than exhibiting uniform flow, the vortices clearly have
preferred channels to move in.  The braided vortex river 
resembles networks of interconnected channels formed by water flowing
over non-cohesive sediment.  Such braided fluvial systems have been
observed from aerial photographs to exist for many different length
scales and types of sediment \cite{braidedrivers,sfgres1,sfgres2}.  In
fact, braiding has been proposed to be the fundamental instability of
laterally unconstrained free surface flow over cohesionless beds, and
has been found to be a robust feature in simulations of river flow
with sediment transport that includes both erosion and redeposition
\cite{naturepaper}. 

\begin{figure}[t] 
\vspace{10cm} 
\begin{center}
    \leavevmode    
\caption[]{ A
``time-lapsed'' photograph of vortex motion with average flow from
left to right.  The lattice size is $600 \times 500$.  Sites
containing an average amount of flow are shown in red.  Yellow sites
have a flow level greater than 20 times the average.  Dark blue sites
have almost no vortex flow, although virtually every site has some
minimal amount.  The intricate braiding pattern is remarkably similar
to the pattern formed by braided fluvial rivers.}
  \end{center}
\end{figure}

 A quantitative scaling analysis reveals that the vortex river pattern
is a self-affine multifractal with scaling dimensions close to those
measured for a variety of braided rivers \cite{bassler2}.  Given the
vastly different length scales and materials involved, this apparent
universality may seem surprising.  Nevertheless, it is known that this
type of universality can exist in systems which evolve by avalanches
into a self-organized critical state \cite{pb}. In the case of braided
vortex rivers, the patterns are due to a slip-stick process consisting
of vortex avalanches, that self-organizes to a critical state
resulting in the observed long-range correlations of the braided
pattern.  It has been postulated that braiding of fluvial rivers is
due to a self-organized critical process\cite{minnesota}.

Are there avalanches in fluvial rivers that could self-organize and
produce the observed braiding?  In fact there are.  ``Pulses'' in
bedload transport have been observed to occur on all spatial and
temporal scales up to those limited by the size of the river studied
\cite{gomez}.  Analogous pulses in the vortex model are seen by
measuring the vortex flow through individual lattice cells as a
function of time.  The flow in a small region of the system is
temporally intermittent; there is a broad distribution of intervals
between pulses, and the pulses themselves can have a broad range of
sizes.  These pulses are a consequence of avalanche dynamics in a
self-organized critical state in the model. Thus, vortices of magnetic
flux are analogous to sediment in fluvial rivers.  The elementary
stick-slip process is that of sediment slipping and then resticking at
some other point, like intermittently moving vortices.  The
elementary slip event can dislodge nearby sediment leading to a chain
reaction of slip events, or avalanches.  Sediment transport can be
triggered when the local sediment slope is too high; the same is true
for vortices in a superconductor.  Thus, in both magnetic flux and
fluvial rivers it appears that the braiding emerges from a stick-slip
process consisting of avalanches of all sizes \cite{bassler2}.

\section{Is Life a Self-organized Critical Phenomenon?}
Evolution has taken place in a highly intermittent way. Periods with
little activity have been pierced by major extinction events where
many species disappeared, and other species emerged. About 50 million
years ago the dinosaurs vanished during such an event, but this is far
from the biggest. 200 million years ago we had the Permian mass
extinction, and 500 million years ago the Cambrian explosion took
place.

Traditional scientific thinking is linear. Nothing happens without
a reason. The bigger the impact, the stronger the response. Thus,
without further ado paleontologists and other scientists working
on early life took it for granted that those extinctions were
caused by some external cataclysmic events.   Several have been suggested,
including climatic changes and volcanic eruptions. The prevailing view on
the Cretaceous event is that it was caused by a meteorite hitting
earth. 

The linear point of view is 
correct for a simple system near equilibrium, 
such as a pendulum nearly at rest. But we do know that large events can 
happen without external impact in geophysical and astrophysical processes.
No meteorite is needed in order to have large earthquakes, for
instance. Actually, there is some striking statistical regularities 
indicating that the mass extinctions are part of a self-organized
critical process.

Species do not evolve in isolation, so biology is a cooperative 
phenomenon! The environment of each individual is made up of other 
individuals. The atmosphere that we breathe is of biological origin, 
with an oxygen content very different from that at the time of the 
primordial soup. Species interact in food webs. 
The interaction can be through competition for resources,
as parasites, or by symbiosis. This allows for the possibility that
the extinction events can be viewed as co-evolutionary avalanches,
where the death of one species causes the death (and birth) of
other species, just as the toppling of one grain of rice in the
rice pile leads to toppling of other grains.

Let us take a look at the fossil record. 
Fortunately, Jack Sepkoski has devoted a monumental effort to mapping out 
the rate of extinction during the last 500 million years.\cite{Sepkoski}$\;$ 
It is extremely important to have as much data as possible, 
since we cannot make accurate theories for specific events, 
and therefore must confront theories with observations at the statistical level.
The insert in Fig.~3 shows the temporal variations of the number of 
Ammonoida families. If part of the curve shown is enlarged, 
the pattern seen on the finer scale
looks the same as that seen on the coarser scale.\cite{SMBB}$\;$  
Thus, there is no typical scale for the variations. 
This {\em scale-independent\/} or {\em self-similar\/} behavior 
is a strong indication of criticality---it cannot occur in simple systems 
with few components, 
including those exhibiting low-dimensional chaotic behavior. 

\begin{figure}[t] 
  \begin{center}
    \leavevmode    
    \psfig{figure=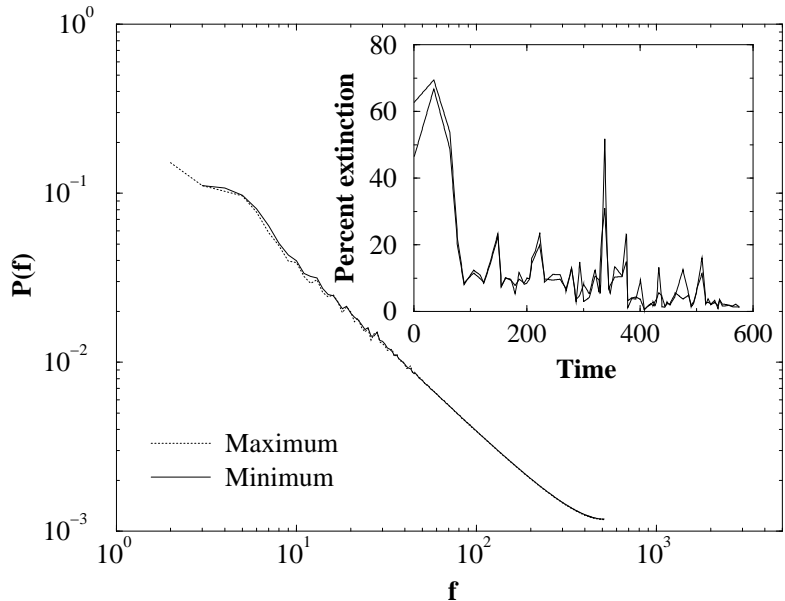,width=8cm,angle=270}
    \caption[]{Spectrum of the
extinction rate for Ammonoida families. The temporal variations of the number
of families  is shown in the insert (Sole, Monrubia, Benton, Bak \cite{SMBB}).}
  \end{center}
\end{figure}

Self-similarity, or scaling, can be expressed more quantitatively in terms
of the power spectrum $p(f)$ of the time series. 
The power spectrum is the Fourier Transform of the autocorrelation function. 
When plotted with log-log axis, Fig.~3,    
it shows an approximately straight line over a couple of decades. 
This indicates that the spectrum is a power law, $p(f) = f^{-\alpha}$. 
The slope $\alpha$ is approximately unity. 
This type of dynamics is called one-over-f ($1/f$) noise. 
It is completely impossible to explain the smooth $1/f$  behavior 
with a set of arguments tailored each to events on a separate scale.
Even in the absence of any theory, the smooth $1/f$ behavior
is an empirical indication that the underlying mechanisms are the same
on all scales.  How else to explain that the curve has the same slope
on all scales, and that segments corresponding to different scales 
join smoothly to form a straight line spanning all scales?
 Figure~4 shows the distribution of life times $T$ of genera, 
also from Sepkoski's data. 
This is another power-law, $N(T) \sim T^{-2}$, 
giving further evidence that life is a critical process.

\begin{figure}[t] 
  \begin{center}
    \leavevmode
\psfig{figure=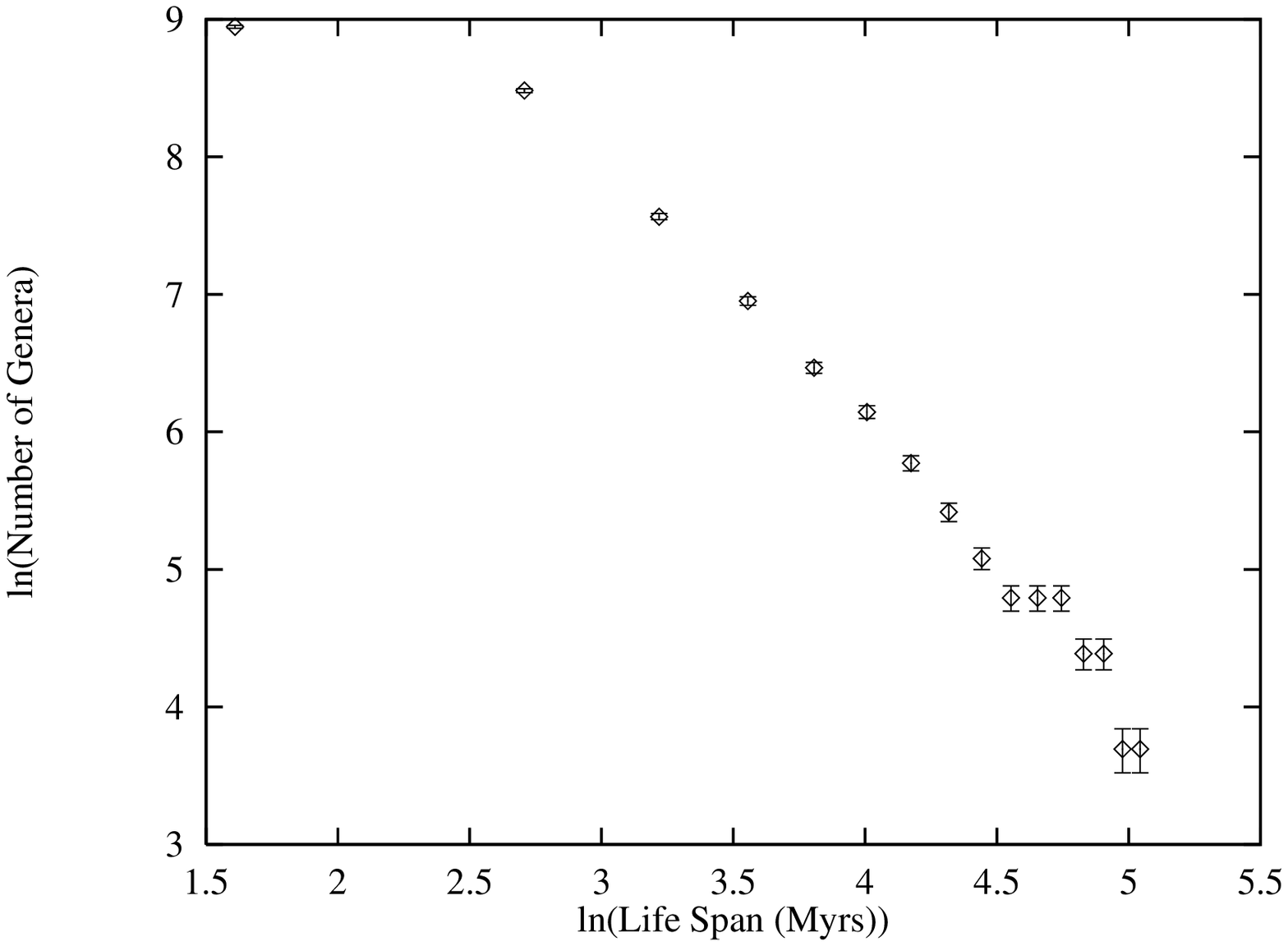,width=8cm,angle=270}
\caption[]{Lifetime distribution for genera as recorded by Sepkoski
and Raup.
The distribution can be well fitted by a power law $N(T) \propto 1/T^{-2}$
except at its lowest $T$-values (Sneppen, Bak, Flyvbjerg, Jensen \cite{PNAS}).}
  \end{center}
\end{figure}

Because of the complexity of the phenomenon that we are dealing with%
---the global biological evomlution on all time scales---mathematical
modelling is an extremely delicate affair. 
It is difficult to go from micro-evolution 
where the mechanisms (genetics) are relatively well understood, 
to macro-evolution at the largest scale. 
Geneticists may understand what goes on within a few generations
of a few hundreds or a few thousands of rats, 
but they have little to say about the behavior 
of an evolving global ecology of millions of species, 
each with hundreds of millions of individuals. 

Kauffman and Johnsen~\cite{K+J} were the first to suggest 
that the Darwinian dynamics of an ecological network 
with all species connected through their interactions,
positive or negative, could lead to a critical state.
The first model for evolution to show SOC 
was the Bak-Sneppen (BS) model \cite{BS,PNAS,pnas}. 

The Bak-Sneppen model represents an entire species by a single fitness number. 
Selection acts on the level of the individual, of course,
but to achieve simplification 
we consider the evolution at the  ``coarse-grained'' species level. 
Consider a number, $N$, of species placed on a circle. 
Each species interacts with its two neighbors. 
Each species is assigned a random fitness $0<f<1$
which represents its ability to survive in a given environment. 
Time is discrete, 
and at each time step  the species with the lowest fitness goes extinct, 
and is replaced by another species with a random fitness $f$, $0<f<1$. 
Alternatively, one could view the process as a pseudo extinction 
where a species is replaced by a mutated variant. 
Whatever the view, 
this change in one species affects the fitnesses of its two neighbors: 
their fitnesses, which might originally have been high, 
are also replaced by new random fitnesses,
reflecting the fact that their   
existence has become a new ball game. 
This process of changing the fitnesses of the least fit species and 
the two it interacts with is continued ad infinitum.

Most of the species have fitnesses above a threshold 
that has established itself with value approximately 0.67, forming
a rather stable network (Figure 5). 
However, there is a localized region with species of lower fitnesses. 
These are the species, or niches, 
that are currently undergoing changes or extinctions as part of a 
{\em co-evolutionary avalanche}.

\begin{figure}[t] 
  \begin{center}
    \leavevmode    
    \psfig{figure=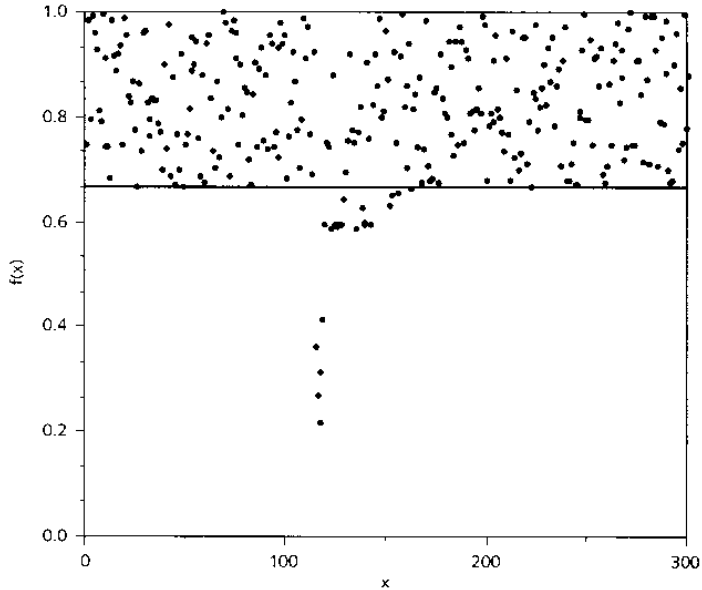,width=10.0cm}
    \caption[]{Illustration of the BS model for 300 species.
    A snapshot of the 300 fitness values is shown.  
    Most values are above the threshold of 0.67.
    The species with fitnesses below 0.67 participate in an avalanche.
    In the next step, the species with the lowest barrier,
    here number 113, will evolve, together with its two nearest neighbors, 
    nos.~112 and 114 (Sneppen, Bak, Flyvbjerg, Jensen \cite{PNAS}).}
  \end{center}
\end{figure}

During an avalanche, 
nature ``experiments'' with the species involved,
changing many of them several times,
until they all have achieved fitnesses above the threshold.
If the changes experienced by any given species is measured vs.\ time, 
one finds punctuated equilibrium behavior, 
with periods of stasis interrupted by intermittent bursts. 
This can be characterized by the power-spectrum of the local activity,
which is a $1/f$ spectrum with exponent $\alpha\sim 0.59$.\cite{pmb}

Note that in the BS model 
evolution progresses by elimination of the least fit species,
and not by propagation of strong species. 
This distinction is not merely semantics.
One can not have a process of evolution, where the individual
species out-competes their environment, the popular view of
Darwinian evolution. The complexity of Life is intimately related
to the existence of large interactive networks. 
Actually, extremal dynamics associated with removing the weakest link is
essential for the emergence of complex or critical phenomena.
The criticality of the SOC earthquake models 
can also be traced to the breakdown of the weakest site, 
and not an arbitrary site. 

Thus, the mechanism of evolution is 
``extinction of the least fit'' rather than ``survival of the fittest''! 
The best a species can hope for is to be a participant of the global
ecological network. In the final analysis, being fit simply means 
being a self-consistent part of a complex structure.

\subsection {Ecology dynamics}

Perhaps the dynamics of evolution can found in a smaller scale
by studying local ecologies or food webs. Keitt and Marquet~\cite{Keitt} 
have studied the dynamics of birds introduced into Hawaiian islands. 
They measured the extinction rate between successive periods of 10 years, 
(to be compared with 4 million year intervals used for the analysis of the
fossil record) and found a power law distribution and  also extracted the 
lifetime distribution of species, yielding another power law with exponent 
near unity. A total of 59 extinctions on six islands were included in their 
statistics. Because of the scant amount of data available, no firm conclusions
could be reached, but everything was consistent with an ecology operating at 
criticality. In a very comprehensive study, Lockwood and Lockwood~\cite{LL} 
have analyzed grasshopper infestations in several regions of Idaho and 
Wyoming. Histograms of annual infestations, measured as the area involved, 
shows a power law distribution. Although numerous external factors affect the 
infestation rate, the results suggest criticality. 

\section{Traffic Jams and the Most Efficient State}

Our everyday experience with traffic jams is that they are annoying
and worth avoiding.  Intuitively, many people believe that if we could
somehow get rid of jams then traffic would be more efficient with
higher throughput.  However, this is not necessarily true.  By
studying a simple model of highway traffic, it is found that the state
with the highest throughput is a critical state with traffic jams of
all sizes.  If the density of cars were lower, the highway would be
underutilized; on the other hand, if it were higher there would
inevitably be a huge jam lowering throughput.  This leaves us with the
critical state as the most efficient state that can be achieved.
Finding a real traffic network operating at or near peak efficiency
may seem highly unlikely.  To the contrary,  as found in the
model, an open network
self-organizes to the critical state \cite{np95}. 

The Nagel-Schreckenberg~\cite{ns92}
model  is defined on a one dimensional lattice with cars moving to
the right.  Cars can move with integer velocities in
the interval $[0,v_{max}]$. 
The maximum velocity $v_{max}$ is typically set equal to 5.
This velocity defines how many ``car lengths'' each car will move
at the next time step.  If a car is moving too fast, it must slow
down to avoid a crash.  A slow moving car
will accelerate, in a sluggish way, when given an opportunity.
The ability to accelerate is slower than the ability to
break.  Also, cars moving at maximum velocity may slow down
for no reason, with
probability $p_{free}$.
A ``cruise-control'' limit of the model exists where $p_{free} \rightarrow
0$.  This means that
all cars which have reached maximum velocity, and have enough headway
in front of them to avoid crashes, will continue to move at maximum
velocity.  Thus it is possible for the motion in the system to be
completely deterministic.

If the cars are moving on a ring starting from random initial
conditions, at low densities the initial jams will ``heal'' and
the system will reach a deterministic state where the current
is equal to the density of cars multiplied by the maximum velocity.
This will hold up to some maximum density above which jams never
disappear and the current is a decreasing function of density.

Remarkably, maximum throughput, $j_{max}$, is
selected automatically when the left boundary condition is an
infinitely large jam and the right boundary is open.\cite{np95} Traffic 
which emerges from the megajam operates
precisely at highest efficiency.  This situation is shown in Fig.~6.

\begin{figure}[t]      
 \begin{center}
    \leavevmode    
\psfig{figure=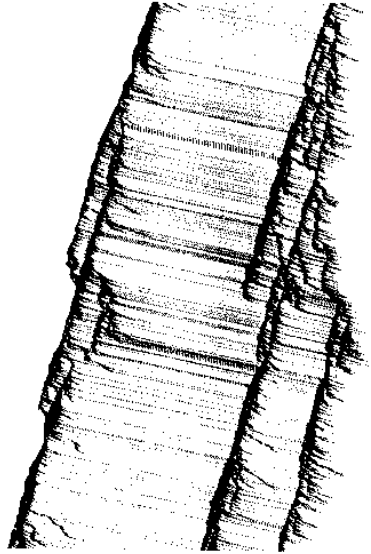,height=3.0in}
\caption[]{Traffic jams. The horizontal direction indicates a highway. Cars 
are shown as black dots. Time progresses in the downward direction. The dots 
form trajectories of individual card. The dark areas with a high density of 
cars indicate traffic jams. Note that the jams are moving backwards.
(Nagel and Paczuski, 1995 \cite{np95}).}
  \end{center}
\end{figure} 

The horizontal axis is space and the vertical axis (down) is
increasing time.  The cars are shown as black dots which move to the
right.  The diagram allows us to follow the pattern in space and time
of the traffic.  Traffic jams show up as dense regions which drift to
the left, against the flow of traffic.  The structure on the left hand
side is the front of the megajam (cars inside the megajam are not
plotted).  Cars emerge from the big jam in a jerky way, before they
reach a smooth outgoing pattern operating at $j_{max}$.  Far away from
the front of the megajam all cars eventually reach maximum velocity.

 If the outflow is perturbed slightly, traffic jams of all sizes
occur.  No cataclysmic triggering event, like a traffic accident, is
needed to initiate large jams.  They arise from the same dynamical
mechanism as small jams and are a manifestation of the criticality of
the outflow regime.  Our natural intuition that large events come from
large disturbances is violated. It does not make any sense to look for
reasons for the large jams. The large jams are fractal, with small
sub-jams inside big jams ad infinitum.  Between the subjams are
``holes'' of all sizes where cars move at maximum velocity.  This
represents the irritating slow and go driving pattern that we are all
familiar with in congested traffic.  On the diagram, it is possible to
trace the individual cars and observe this intermittent pattern.  This
behavior gives rise to $1/f$ noise, as seen in real traffic flow
\cite{musha}.  This $1/f$ behavior can be calculated exactly for this
model by formulating the jams as a cascade process \cite{np95}.  The
picture of avalanche dynamics as a fractal in space and time has
application to many complex dynamical systems in addition to
traffic.\cite{pmb}

The conventional view is that one should try to get rid of traffic
jams in order to increase efficiency and productivity.  However, the
critical state, with traffic jams of all sizes, is the most efficient
state that can actually be achieved.  A carefully prepared state where
all cars move at maximum velocity would have higher throughput, but it
would be dreadfully unstable.  The very efficient state would
catastrophically collapse from any small fluctuation.  A similar
situation occurs in the familiar sand pile models of SOC.\cite{BTW}
One can prepare a sand pile with a supercritical slope, but that state
is unstable to small perturbations.  Disturbing a supercritical pile
will cause a collapse of the entire system in one gigantic avalanche.

But there is perhaps even a deeper relationship between traffic and
economics \cite{np95,B}. In an economy, humans interact by
exchanging goods and services.  In the real world, each agent has
limited choices, and a limited capability to monitor his changing
environment.  This is referred to as bounded rationality.  The
situation of a car driver in traffic can be viewed as a simple example
of an agent trying to better his condition in an economy.  Each
driver's maximum speed is limited by the other cars on the road and
posted speed limits.  His distance to the car in front of him is
limited by his ability to stop and his need for safety in view of the
unpredictability of other drivers.  He is also exposed to random
shocks from the road or from his car.  He may be absent minded.  If
traffic is a paradigm for economics in general, then perhaps we have
found a new economic principle: the most efficient state that can be
achieved for an economy is a critical state with fluctuations of all
sizes.

\section{The Critical Brain}

Why do we need a brain at all? In a sub-critical world everything would
be simple and uniform - there would be nothing to learn. In a supercritical
world, everything would be changing all the time in a chaotic way -
it would be impossible to learn. The brain is necessary for us in order
to navigate  in a complex, critical world.

A brain is able not only to remember, but also to forget and adapt
to a new situation. In a sub-critical brain memories would be frozen.
In a supercritical brain, the patterns would change all the time so 
no long term memory would be possible. This leaves us with one choice -
the brain itself has to be in the in-between critical state. Using  physics
terminology, it is the high susceptibility  of the critical state which
makes it adaptable.

Actually,  Alan Turing \cite{Turing}, some time ago,
 speculated that perhaps the 
working brain needs to operate at a barely critical level, in order
to stay away from the two extremes - namely the too correlated
sub-critical level, and the too explosive supercritical
dynamics.

In traditional neural network models, the goal has typically been to
have the desired patterns represented by very stable states. In the
Hopfield model \cite{Hopfield}, for instance, the patterns correspond
to deep energy minima in a spin glass model. This represents the
traditional Hebbian \cite{Heb} picture where synapses connecting
firing neurons are strengthened. Once the desired memory has been
encoded, it is hard to adapt to a new situation when the environment
changes, because the deep minima have to be removed by a dynamical
process.  Traditional models are sub-critical. Moreover, the learning
process takes place by having an external teacher, a computer
algorithm that sets the strengths of the neural network connections.
It is hard to see how this can be accomplished without the
intervention of an external agent. The learning process of the
neural network is not
self-organized. Chialvo and Bak \cite{Dante} have suggested an
alternative scheme, which at least in principle could act as a
paradigm for real brain processes.

\subsection{Learning from Mistakes}\label{subsec:wpp}
Recall that in the evolution model,  criticality, and hence
 complexity,  was achieved by extremal dynamics where the
least fit species were weeded out. Chialvo and Bak
used a similar mechanism for brain functioning, with the synapses
playing the role of the individual species. Whenever a poor
result is achieved, all the synapses which fired in the process
are democratically
punished.  However, good behavior is not rewarded at all;
the reward system is all stick and no carrot.
There is 
no Hebbian strengthening of successful synapses. While the model
is grossly simplified, the features of the model are all
biologically plausible.

\begin{figure}[htbp]
\centering
\psfig{figure=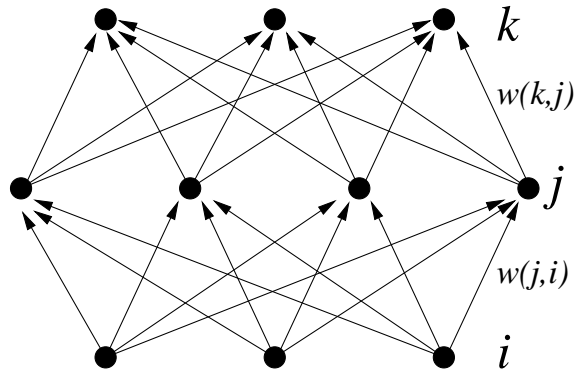,height=3.0in,angle=-90}
\vspace{0.3truein}
\caption{Neurons in layer $J$ have synaptic connections with all
inputs in $I$ and connect with all neurons in layer $K$.}
\end{figure}

The topology of the network is not very important. For simplicity,
let us consider neurons arranged in the layered network in Fig.~7,
where $K$ represents the outputs, $I$ the inputs and $J$ the middle layer. 
Each input is connected with each neuron in the middle layer which,
in turn, is connected with each output neuron, with weights $W$ representing
the synaptic strengths. The network must learn to connect each input with 
the proper output (which is pre-determined) for any arbitrary associative 
mapping. The weights are initially randomised, $0<W<1$.

The dynamical process in its entirety is as follows:\\
An input neuron is chosen. The neuron $j_m$ in the middle layer with the 
largest $w(j,i)$ is firing. Next, the output neuron $k_m$ with the maximum 
$w(k,j_m)$ is firing. If the output $k$ happens
to be the desired one, {\bf  nothing }is done, otherwise
$w(k_m,j_m)$ and $w(j_m,i)$ are both depressed by a fixed amount.
The iterative application of this rule leads to a convergence to any 
arbitrary input-output mapping. Since there are no further changes once
the correct result has been achieved, the proper synapses are only
barely stronger than some of the incorrect synapses.

Supposed now that the environment changes, so that a different connection
between input and output is correct. The neurons which fire and led to
the previously correct output are now punished, allowing new connections.
Eventually that pattern will also quickly be learned.

The reason for quick re-learning (adaptation)
is simple. The rule of adaptation assures that synaptic changes only
occur at neurons involved in wrong outputs. The landscape
of weights is only re-shaped to the point where the new winners barely 
support the new correct output, with the old pattern only slightly
suppressed. Thus, only a slight suppression of a currently active pattern is
needed in order to generate new patterns when need be. In particular, 
re-learning of ``old'' patterns which have been correct once in the past is
fast. This feature can be strengthened if the synapsis which have never
been firing when a good result was achieved are punished more than synapses
whose firing has previously led to a good result. 

The 
landscape of synaptic strength in our model after many learning cycles 
consist of very many values  which are very close to those of the
active ones, a manifestation of the critical nature of the state.
Figure 8 shows a snapshot of the synaptic strengths. The synapse indicated
by an arrow is a currently active one, associated with a correct response.
Other neurons near the active surface have strengths located slightly
below the critical surface.
One can imagine that ``thinking'' is the process of sifting through,
and suppressing, patterns which once have been correct, until 
a combination leading to a good result is achieved. Bits and pieces
of patterns that have previously been successful are utilized.
Old memories are located at the same spot where they have always been -
they have simply been slightly suppressed by more recent patterns.

\begin{figure}[htbp]
\vspace{-.5truein}
\centering
\psfig{figure=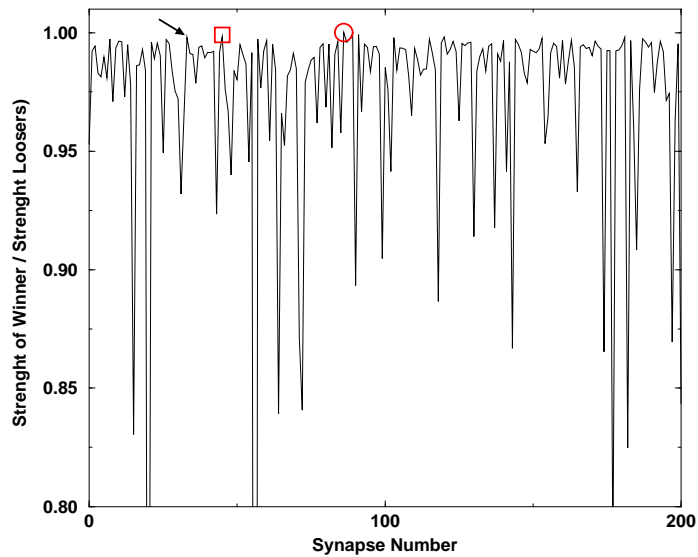,width=3.5in,angle=-90,clip=true}

\caption{\footnotesize{Landscape of synaptic strenghts between
one input and 200 neurons in the middle layer. The encircled value
corresponds to the strongest connection between the input neuron and
the output layer. The arrow and the box indicate other connections
likely to be used in the future (see text). }}
\end{figure}

The biological plausibility of the schema depends on the realization at 
the neuronal level of two crucial features:

a) Activity propagates through the strongest connections, i.e. extremal,
or winner-take all, dynamics. This can be fulfilled by a local circuit 
organisation, known to exist in all cortices, where the firing of
other neurons is shut off by lateral inhibitory connections.

b) Depression of synaptic efficacy involves the entire path of firing
neurons. A process must exist such that punishment can be relayed long after
the neuron has fired, when the response from the outer world to the action
is known. Chialvo and Bak conjectured a mechanism of ``tagging''  synapses 
for subsequent punishment, or long term depression (LTD), analogous to (but 
mirroring) recently reported tagging of synapses for long term potentiation
(LTP) \cite{tagging}. The feed-back probably takes place through the limbic
system of neurons, situated in the neck, which spray the large areas of
the brain. One could imagine that this global feed-back signal
affects all neurons which have recently fired, causing plastic changes
of the synaptic connections. The limbic system is disconnected when
dreaming, which could explain why we generally do not remember our dreams.
Actually, long-distance, long-term
synaptic depression has been directly demonstrated by Fitzsimons et al
\cite{Fitz} in cultured hippocampal neurons from rat embryos. 

In addition to giving insight into mechanisms for learning in the brain,
the ideas presented here could be useful for artificial learning 
processes, for instance in adaptable robots. These possibilities are 
currently being investigated and appear promising. 

Historically, many
processes that were considered to be examples
 of  directed learning have 
been shown to be caused by selection. The Larmarquean theory of 
evolution as a learning process, where useful acquired features are
strengthened, was replaced by the Darwinian theory of evolution as a selection 
process, where the unfit species are weeded out. A similar paradigm shift 
occurred in immunology through the theory of clonal selection. Ironically, if 
the philosophy represented by the Chialvo-Bak model is correct,
learning in the brain is not a (directed) learning process either.
It is  also an example of a
co-evolutionary selection process where incorrect connections are weakened.

The paradigm of science in the second millenium,
reductionism, is insufficient to explain complexity in nature. There 
appears to be a need for
an outside organizing agent who fine tunes the natural world and puts 
the building
blocks together. We speculate that, instead of this agent,
co-evolutionary selection leading to a critical state
by removing untenable parts
may be the fundamental organizing
principle leading to all the possible complexity
in the universe. 

\eject
\end{document}